# Manganese reduction/oxidation reaction on graphene composites as a reversible process for storing enormous energy at a fast rate

Yanyi Chen[1,2], Chengjun Xu[1]*, Shan Shi[1,2], Jia Li[1], Feiyu Kang[1,2], Chunguang Wei[3]

[1]Graduate School at Shenzhen, Tsinghua University, Shenzhen City, Guangdong Province, 518055, P. R. China
[2]State Key Laboratory of New Ceramics and Fine Processing, School of Materials Science and Engineering, Tsinghua University, Beijing, 100084, P. R. China
[3]Cubic-Science Co. Ltd., Shenzhen City, Guangdong Province, 518055, P. R. China
Corresponding author: Chengjun Xu, E-mail: vivaxuchengjun@163.com

**Abstract:** Oxygen reduction/evolution reaction (ORR/OER) is a basic process for fuel cells or metal air batteries. However, ORR/OER generally requires noble metal catalysts and suffers from low solubility ($10^{-3}$ molar per liter) of $O_2$, low kinetics rate ($10^{-6}$ cm$^2$/s) and low reversibility. We report a manganese reduction/oxidation reaction (MRR/MOR) on graphene/$MnO_2$ composites, delivering a high capacity (4200 mAh/g), fast kinetics ($2.4\times10^{-3}$ cm$^2$/s, three orders higher than ORR/OER), high solubility (three orders than $O_2$), and high reversibility (100%). We further use MRR/MOR to invent a rechargeable manganese ion battery (MIB), which delivers an energy density of 1200 Wh/Kg (several times of lithium ion battery), a fast charge ability (3 minutes), and a long cycle life (10,000 cycles). MRR/MOR renders a new class of energy conversion or storage systems with a very high energy density enabling electric vehicles run much more miles at one charge.

Oxygen reduction/evolution reaction (ORR/OER) occurring on an air electrode is a basic process for energy conversion or storage. Electrochemical energy conversion systems, for example full cells, or storage systems, such as metal (Li, Al, Zn, Na, *etc*.) air batteries, generally use air cathode to obtain a high capacity from ORR/OER. However, ORR/OER generally requires noble metal catalysts and suffers from low kinetics rate ($10^{-6}$ cm$^2$/s), low solubility ($10^{-3}$ molar per liter) of $O_2$ in electrolyte, low reversibility, and drought of electrolyte[1,2]. Here we report a reversible manganese reduction/oxidation reaction (MRR/MOR) process between $Mn^{4+}/Mn^{2+}$ potentially for conversing or storing enormous energy. MRR/MOR, occurring on graphene/$MnO_2$ composites, delivers an anomalous high capacity (4200 mAh/g), a fast kinetics ($2.4\times10^{-3}$ cm$^2$/s, three orders higher than ORR), a high solubility (three orders than $O_2$), and high reversibility (100%). We further use MRR/MOR to invent a manganese ion battery (MIB), which uses graphene/$MnO_2$ cathode, Zn cathode, and the aqueous electrolyte containing $Zn^{2+}$ and $Mn^{2+}$ ions. The MIBs deliver a high energy density (1200 Wh/Kg, several times than lithium ion batteries used in Tesla Model S), a fast charge ability (3 minutes), and a long cycle life (10,000 cycles). Our results indicate that MRR/MOR renders a new class of energy conversion or storage systems and the new MIBs open possibility that the electric vehicles run much more miles at one charge.

Oxygen reduction/evolution reaction (ORR/OER) occurring on an air electrode is a worldwide interesting and basic process for conversing or storing renewable energy. The ORR converts chemical energy directly into electrical energy resulting in a high capacity. ORR on an electrode involves in the following steps. $O_2$ is firstly solubilized in the electrolyte (aqueous or organic solutions). Then it diffuses in the electrolyte, is adsorbed on the surface, and is reduced to oxides on the interface of three phases (solid, water and gas). Electrochemical energy conversion systems, for example full cells, or storage systems, such as metal air batteries, generally use air electrode to obtain a high capacity from ORR. The air electrode generally consists of porous support materials, for example carbons, oxides, metals, *etc*., on which ORR or OER occurs. The catalysts, for example noble metals (Pt, Ru, Pd), are frequently loaded on the porous support materials for improving the kinetics of ORR or OER.

Due to the energy crisis, the society requires rechargeable batteries with high capacity and long cycle life to power the portable electronics for further long time, to drive the electric vehicles rivaling cars powered by the combustion engine, and to store electricity generated by renewable sources. Lithium ion batteries (Li-ion) now are predominating in portable electronics and in electrification of transport (for example in Tesla Model S). However, the energy density of Li-ion batteries is insufficient for the long-term needs of society, for example, long-range (over 500 miles) electric vehicles. In addition, the safety issue of lithium ion batteries is an inevitable problem. It is a formidable task to go beyond the lithium ion battery chemistry for the high capacity and safety electrochemical energy systems. The current worldwide interest is in high energy density conversion or storage systems based on ORR/OER, for example fuel cells or lithium air (Li-$O_2$) batteries. The typical ORR in the proton fuel cell is written as:

$$O_2 + 2H_2O + 4e^- \rightarrow 4OH^- \qquad (1)$$

While the ORR/OER in the rechargeable Li-$O_2$ batteries is

$$O_2 + 2Li^+ + 2e^- \leftrightarrow Li_2O_2 \qquad (2)$$

The ORR/OER delivers a high capacity in fuel cells or in metal air batteries, leading to a high energy density, several times higher than lithium ion batteries. However, ORR/OER generally requires noble metal catalysts and suffers from low solubility ($10^{-3}$ molar per liter) of $O_2$ in electrolyte, low kinetics rate ($10^{-6}$ cm$^2$/s) and drought of electrolyte due to the open air electrode. These drawbacks hinder the electrochemical systems based on ORR/OER to deliver the energy at a fast rate. In addition, the open electrode structure to air needs complicated technologies and high cost to prevent the evaporation of the electrolyte. Furthermore, the reversibility of ORR/OER is very low so that most of metal air batteries are primary. And the cycle life of secondary metal air batteries is very short (several hundred cycles).

Despite of oxygen, manganese exists in various valence states, for example $Mn^{2+}$, $Mn^{3+}$, $Mn^{4+}$, $Mn^{6+}$, or $Mn^{7+}$, in which $Mn^{4+}$ exists as solid $MnO_2$ and $Mn^{2+}$ ions is highly soluble in aqueous solutions up to 2 molar per liter (three orders higher than $O_2$ in aqueous or organic electrolytes), respectively. A typical electrochemical manganese reduction/oxidation reaction (MRR/MOR) is written as:

$$Mn^{2+} + 2H_2O \leftrightarrow MnO_2 + 4H^+ + 2e^- \quad (3)$$

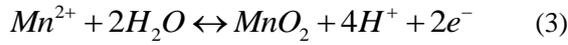

The MOR from $Mn^{2+}$ ions in the aqueous solution to solid $MnO_2$ deposit is frequently used to electrodeposite $MnO_2$ for various applications. However, the MRR/MOR as written in *equation 3* has never been used as a basic reversible process, similar to ORR/OER, for conversing or storing enormous energy. This is because it is hard to discover porous materials with certain physical and electronic structures, which supports and favorites the MRR/MOR with a high reversibility.

Graphene is several layers of carbon atoms arranged in a hexagonal lattice with two carbon atoms per unit cell. Of the four valence states, three $sp^2$ orbitals form a $\sigma$ state with three neighboring carbon atoms, and one $p$ orbital develops into delocalized $\pi$ and $\pi^*$ states that are degenerated at the corner of the hexagonal Brillouin zone. This degeneracy occurs at the Dirac crossing energy resulting in a pointlike metallic Fermi surface. The unique electronic structure and large surface favorites the charge transfer. $Li-O_2$ or lithium sulfur (Li-S) batteries use graphene as the cathode, where supports and favorites the oxidation/reduction of oxygen or sulfur with lithium ions to trigger the formation/decomposition of $Li_2O_2$ or of $Li_2S_2$, respectively. Therefore, graphene is possible to enable MRR/MOR reversible.

We firstly investigated the reversibility of MRR/MOR on graphene in the aqueous $MnSO_4$ solution. The cyclic voltammetry (CV) as shown in Figure 1a clearly shows the MRR/MOR. The MOR from soluble $Mn^{2+}$ ions to $MnO_2$ deposits occurs at 1.58 V vs. $Zn^{2+}/Zn$, while the MRR from $MnO_2$ to soluble $Mn^{2+}$ ions follows the typical process of two electrons occurring at 1.35 V and 1.20 V vs. $Zn^{2+}/Zn$, respectively. The MOR is widely adopted to be used to synthesize $MnO_2$ materials on different substrates. However, the MRR from $MnO_2$ to soluble $Mn^{2+}$ ions has never been investigated to supply the electrons. As shown in Figures 1a and 1b MOR can store electric energy (electrons) by solidifying the $Mn^{2+}$ ions to $MnO_2$ deposits, while the MRR releases the stored electric energy by dissolving $MnO_2$ solid to soluble $Mn^{2+}$ ions. The kinetic rate ($D_0$) of the MRR/MOR can be calculated by $i_p = 2.69 \times 10^5 n^{3/2} A D_0^{1/2} C^* v^{1/2}$, where $i_p$ is the peak current, $n$ is electron transfer number, $A$ is the area of graphene, $C^*$ is the initial concentration of $Mn^{2+}$ ions, and $v$ is the sweep rate. The kinetic rate ($D_0$) is $2.4 \times 10^{-3}$ cm$^2$/s, which is three order higher than the ORR/OER of $10^{-6}$ cm$^2$/s. We use galvanic charge/discharge method to investigate the capacity and the reversibility of the MRR/MOR. The charge/discharge curve of graphene electrode at the voltage ranging from 1.00 V to 1.90 V vs. $Zn^{2+}/Zn$ is shown in Figure 1b. The charge capacity is about 4700 mAh g$^{-1}$, while the maximum discharge capacity is about 4200 mAh g$^{-1}$ at a current density of 0.1 Ampere per gram (A g$^{-1}$). The reversibility between MOR and MRR is 89.4%, which is higher than the reversibility of ORR/OER[3,4].

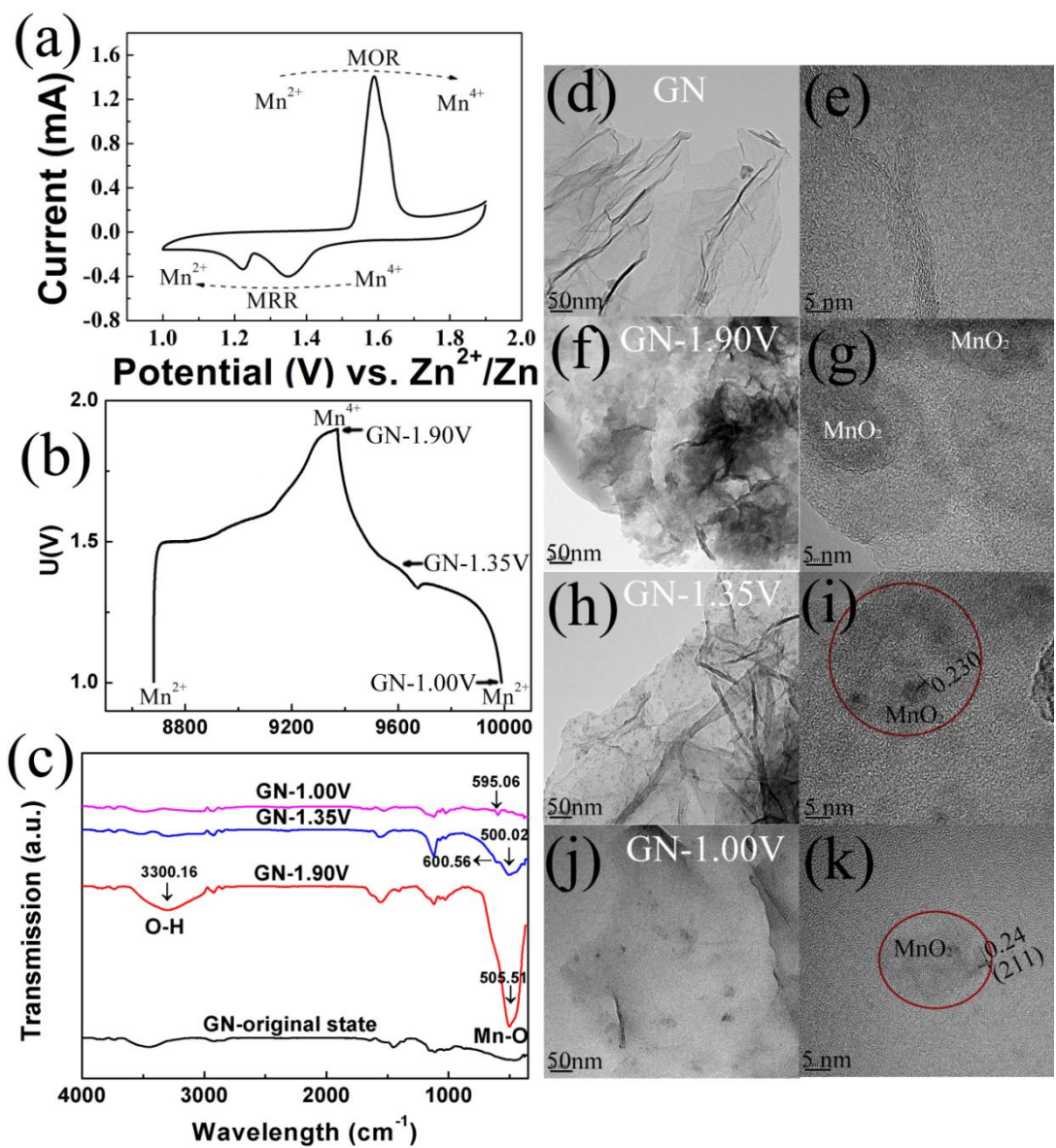

Figure 1 (a) Cyclic voltammetry curve of manganese oxidation/reduction reaction (MOR/MRR). (b) Galvanic charge/discharge curve of MOR/MRR. (c) FTIR curves of fresh graphene electrode (GN-original state) and graphene electrodes at full charge state (GN-1.90V), intermediate discharge state (GN-1.35V), and full discharge state (GN-1.00V), respectively. (d) TEM image of graphene (GN). (e) HRTEM image of graphene. (f) TEM image of graphene at full charge state (GN-1.90V). (g) HRTEM image of graphene at full charge state (GN-1.90V). (h) TEM image of graphene at intermediate discharge state (GN-1.35V). (i) HRTEM image of graphene at intermediate discharge state (GN-1.35V). (j) TEM image of graphene at full discharge state (GN-1.00V). (k) HRTEM image of graphene at full discharge state (GN-1.00V).

We further investigate the process of MRR/MOR. The fresh graphene electrode is charged and discharged after ten cycles, charged to 1.90 V and then discharged to 1.35 V and 1.00 V, respectively. The graphene at full charge state of 1.90 V is denoted as GN-1.90V and the graphenes discharged to 1.35 and 1.00 V are denoted as GN-1.35V and GN-1.00V, respectively. The transmission electron microscope (TEM) and high resolution transmission electron microscope (HRTEM) images of fresh graphene (GN) are shown in Figure 1d and 1e, respectively. The graphene consists of several layers of carbon atoms and the surface of graphene is smooth and clean. However, at the full charge state of 1.90 V (GN-1.90V), there are sheet deposits covering the entire graphene surface as shown in Figures 1f and 1g. The sheet deposits are confirmed as $MnO_2$ and the valence state of Mn of deposits is confirmed to be four by TEM mapping and X-ray photoelectron spectrum analysis (Figures S4e). $MnO_2$ deposits are formed from $Mn^{2+}$ ions in the aqueous electrolyte during the MOR. The $MnO_2$ deposits show sheet morphology with 50~100 nm in diameter and 8.3 nm in thickness, which is measured by atomic force microscope (Figure S8). When the graphene electrode is discharged to 1.35 V, the coverage of the $MnO_2$ deposits on graphene surface decreases (as shown in Figures 1h and 1i). And the diameter and thickness of $MnO_2$ deposits are deduced to 20 nm and 7.5 nm for GN-1.35V, respectively. When the graphene is further discharged to 1.00 V at full discharge state, there are barely $MnO_2$ residual and the most of graphene surface is smooth and clean again as shown in Figures 1j and 1k. The diameter and thickness of $MnO_2$ residual are further deduced to 10 nm and 4.4 nm, respectively. The TEM measurements on the graphene from full charge state of 1.90 V to full discharge state of 1.00 V clearly confirm the reversible MOR/MRR process between soluble $Mn^{2+}$ ions and $MnO_2$ solid. The FTIR (Figure 1c) and Raman spectra (Figure S4) further confirm the MRR. The FTIR curves of graphene from full charge state of 1.90 V to intermediate discharge state of 1.35 V and to full discharge state of 1.00 V show the gradually weakness of Mn-O stretch peaks at ca. 500 nm. At the full charge state the graphene surface is covered by $MnO_2$ deposits and the stretch strength of Mn-O bond is very strong. However, the stretch strength of Mn-O bond will weaken at 1.35 V and almost vanish at 1.00 V. The physicochemical characterization on graphene electrode from full charge state to full discharge state confirms the occurrence and the high reversibility of MRR/MOR. In addition, for GN-1.35V and GN-1.00V the peak at ca. 600 nm corresponds to the formation of $ZnMn_2O_4$, which confirms the insertion of $Zn^{2+}$ ions into $MnO_2$ during discharging. This result is accordance with the X-ray diffraction pattern (Figure S4b). The MRR/MOR delivers a very high capacity up to 4200 mAh $g^{-1}$, which is potentially used for storing or conversing enormous energy.

To further improve the reversibility of MRR/MOR, we synthesized the graphene/$MnO_2$ composites with the mass content of graphene ranging from 5% to 40%. The graphene/$MnO_2$ composites are denoted as 5%GN to 40%GN, in which the former number represents the mass percentage of graphene. The TEM images of 10%GN, 20%GN and 40% GN are shown in Figures 2a, 2b and 2c. There are $MnO_2$ sheets on the graphene surface. $MnO_2$ sheets are 20 nm in width and 2 nm in thickness. The XRD and FTIR patterns are exhibited in Figures 2d and 2f. The $MnO_2$ in graphene/$MnO_2$ composites shows a similar sheet morphology and crystalline structure (α-MnO$_2$) with the MnO$_2$ deposit on graphene formed by MOR as shown in Figure 1. FTIR curves of graphene composites indicate that with increase in the mass percentage of graphene from 0% to 100%, the stretch of Mn-O bond at 500 nm will gradually weaken and finally vanish. The similar occurrence of weakness of Mn-O bond at ca. 500 nm in FTIR curves of Figure 1c and Figure 2d confirm again the occurrence of the MRR from MnO$_2$ deposits to Mn$^{2+}$ ions. We use galvanic charge/discharge method to investigate the capacity and the reversibility of the MRR/MOR occurring at graphene/MnO$_2$ composites. With the increase of MnO$_2$ content the reversibility of the MRR/MOR will increase to 100%, while the maximum capacity is deduced from 4200 to 1700 mAh g$^{-1}$ at a current density of 0.1 A g$^{-1}$. The presence of MnO$_2$ on graphene in advance provides the extra MnO$_2$ source for MRR, which improves the reversibility of the MOR/MRR, while the extra weight of MnO$_2$ in graphen/MnO$_2$ composites decreases the specific capacity of the composite. However, the most important is that we can manually and arbitrarily adjust the reversibility and capacity of the MRR/MOR, while it can never be done in ORR/OER. This leads to a great advantage of MRR/MOR over ORR/OER.

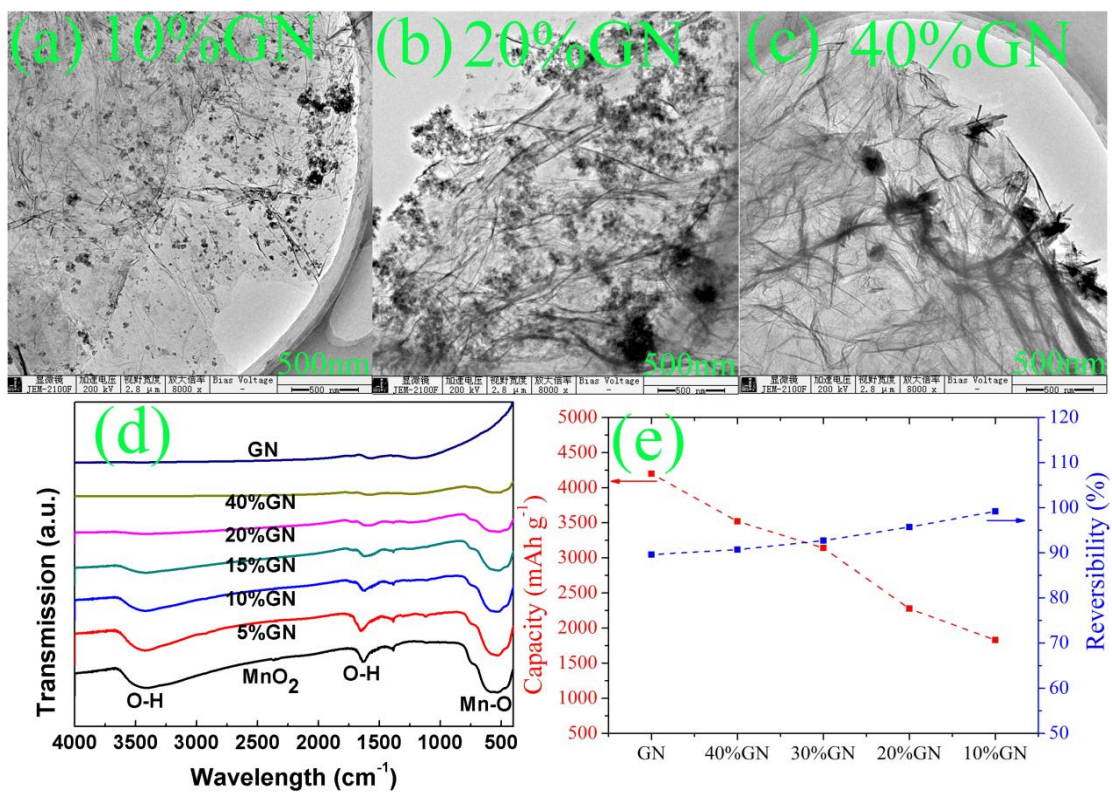

Figure 2 (a) TEM image of 10%GN. (b) TEM image of 20%GN. (c) TEM image of 40%GN. (d) FTIR curves of graphene (GN), graphene/MnO$_2$ composites and MnO$_2$. (e) The capacity and reversibility of manganese oxidation/reduction reaction on graphene (GN) and graphene/MnO$_2$ composites at a current density of 0.1 Ampere per gram.

The measurement of TEM, XRD, FTIR, element mapping, AFM, and Raman spectra clearly proves the reversible MOR/MRR on graphene (or graphene/MnO$_2$ composites). The MRR/MOR delivers a capacity up to 4200 mAh/g, a fast kinetics ($2.4 \times 10^{-3}$ cm²/s), high reversibility (100%), and a high solubility. The MRR/MOR, similar to ORR/OER, shows great potential for conversing or storing enormous energy at a fast rate. In the following, we use the MRR/MOR to invent a multivalent ion battery with ultrahigh energy density and power density.

The recent reports revealed the conception of multivalent ion battery chemistry beyond lithium battery chemistry. The usage of multivalent ion gives us new choices to design rechargeable batteries with desirable properties, for example zinc ion battery[5]. The zinc ion battery generally uses a manganese dioxide cathode, an electrolyte containing $Zn^{2+}$ ions, and Zn anode. Since each multivalent ion being stored in the host materials results in charge storage of over one electron, storage of multivalent ions presents a higher capacity than storage of univalent ions. Therefore, the multivalent ion battery chemistry coupled with new nanostructured host materials has possibility to invent new batteries with a higher capacity to meet our needs. Here, we use MRR/MOR to invent the new graphene-based multivalent ion battery (MIB) chemistry beyond lithium battery chemistries. The structure and battery chemistry of graphene-based MIBs are shown in Figure 3a. Graphene-based MIBs use graphene (or $MnO_2$/graphene composite) as cathode, zinc as anode, an aqueous solution containing $Zn^{2+}$ and $Mn^{2+}$ ions as the electrolyte, respectively. The battery chemistry of the graphene-based MIBs can be written as the cathodic and anodic reactions as shown in equations 4, 5 and 6, respectively. The graphene-based MIB in this report just likes one battery, but consists of a fuel cell (Equations 4 and 6) and an intercalation battery (Equations 5 and 6).

The cathodic reactions are:

$$Mn^{2+} + 2H_2O \leftrightarrow MnO_2 + 4H^+ + 2e^- \qquad (4)$$

$$ZnMn_2O_4 \leftrightarrow Zn^{2+} + 2e^- + 2MnO_2 \qquad (5)$$

The anodic reaction is

$$Zn^{2+} + 2e^- \leftrightarrow Zn \qquad (6)$$

It is shown previously that the reversibility of MRR/MOR on graphene is 89.4%, which indicates that there is $MnO_2$ residual on the surface of graphene. The same result is discovered in metal air batteries for example in Li-$O_2$ batteries, in which $Li_2O_2$ cannot be fully oxidized and the waste $Li_2O_2$ is left. The accumulation of $Li_2O_2$ trash results in the decay of capacity of Li-$O_2$ batteries. However, for MRR/MOR in MIBs there are three advantages over ORR/OER in metal air batteries. The first is that the reversibility of MRR/MOR is much higher than that of ORR/OER, which decreases the accumulation of $MnO_2$. The second is that the reversibility of MRR/MOR can be improved to 100% by adding $Mn^{4+}$ source ($MnO_2$) in advance for example $MnO_2$/graphene composites as cathode. The third is that the $MnO_2$ residual can reversibly store/release $Zn^{2+}$ ions during discharge/charge as shown in *equation 5* to contribute a high capacity of 308 mAh/g. The occurrence of insertion/extraction of $Zn^{2+}$ ions into/from $MnO_2$ residual is confirmed by XRD (Figure S4d), TEM mapping

(Figure S7), FTIR (Figure S4d) and XPS analysis (Figure S4e). The synergistic reaction of MRR/MOR (*equation 4*) and storage/release of $Zn^{2+}$ ions into/from $MnO_2$ residual (*equation 5*) enable the MIBs deliver a very high capacity, high reversibility, and a high stability. And the fast kinetics rate of MRR/MOR enables the MIBs capable of storing and delivering enormous energy at a fast rate.

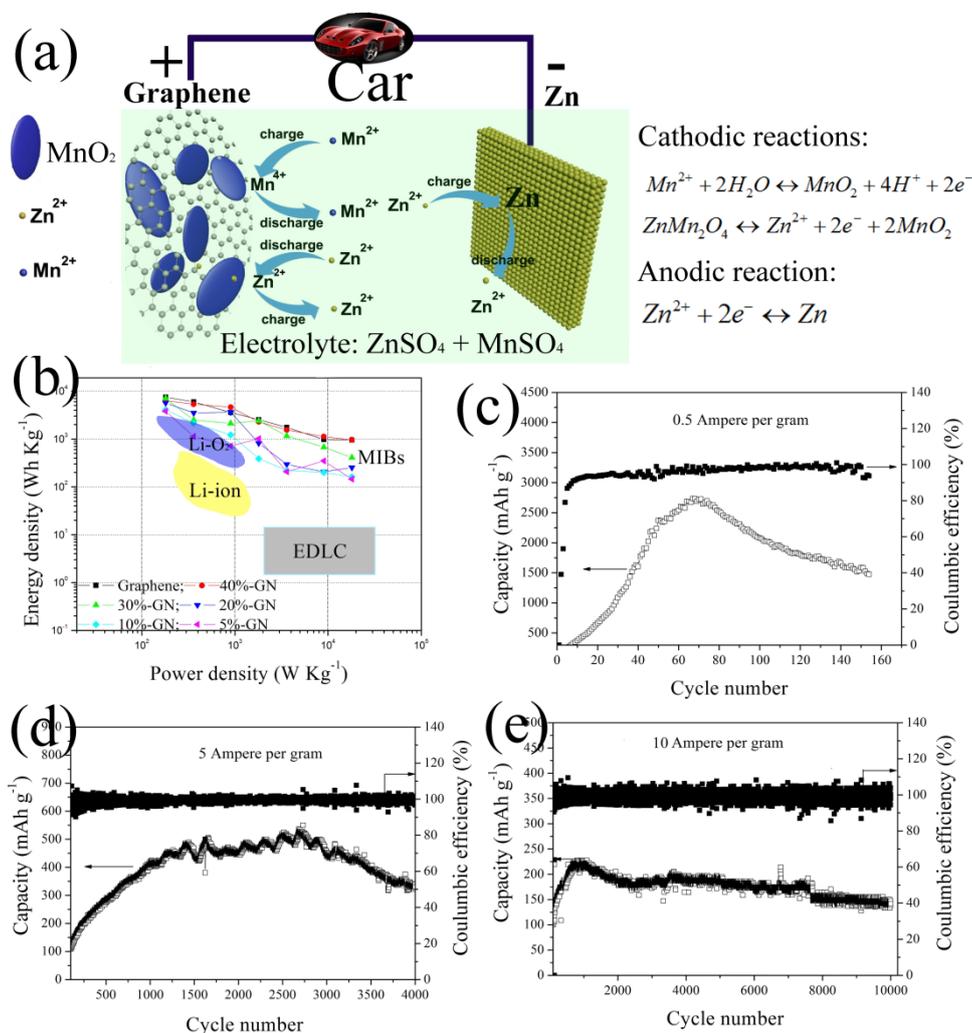

Figure 3 (a) Battery chemistry and structure of graphene-based multivalent ion battery (MIB); (b) Ragone plot of graphene-based MIBs, lithium ion batteries (Li-ion), lithium air batteries (Li-O$_2$) and supercapacitors (EDLC); (c) Cycle life and Coulumbic efficiency of graphene-based MIB at a current density of 0.5 Ampere per gram; (d) Cycle life and Coulumbic efficiency of graphene-based MIB at a current density of 5 Ampere per gram; (e) Cycle life and Coulumbic efficiency of graphene-based MIB at a current density of 10 Ampere per gram.

We have assembled the prototype of MIBs based on the MnO$_2$/grapheme composites (See methods). The charge/discharge curves of MIBs are shown in Figures S14-S20. The Ragone plot of the graphene-based MIBs in terms of energy and power densities as well as the lithium ion batteries (Li-ion), rechargeable Li-O$_2$ batteries, and supercapacitors (EDLC) are shown in Figure 3b. The maximum energy densities of MIBs based on MRR/MOR are up to 5000 ~ 6500 watt hours per kilogram (Wh/kg), which are higher than 3000 ~ 5000 Wh/kg of Li-$O_2$ batteries. If the active material is 5% ~ 15% weight of total battery. The practical energy density of MIBs is up to 1200 Wh/kg, which is 3-5 folds higher than Li-ion used in Tesla Model S (250～400 Wh/kg) [2, 3, 4, 16]. The MIB can be charged and discharged at a current density of 10 A $g^{-1}$, leading to a power density of 1.8 kilowatts per kilogram (KW $Kg^{-1}$), which is one order higher than Li-ion or Li-$O_2$ batteries. In addition, as shown in Figures 3c, 3d and 3e, MIB shows a long cycle life (over 10000 cycles) at a very high current. Furthermore, MIBs are constructed from abundant and eco-friendly materials, such as manganese, the aqueous electrolytes, manganese dioxide, zinc or carbon materials, while lithium is rare. Lithium batteries, for example Li-ion, Li-$O_2$, Li-S, suffer from safety issues due to high activity of lithium, while the materials used in MIBs are inert to the fire. Therefore, MIBs based on the MRR/MOR can be used to power the portable electronics for further long time, to drive the electric vehicles rivaling cars powered by the combustion engine, and to store electricity generated by renewable sources. Our results indicate that the safety and eco-friendly multivalent ion battery chemistry is comparable to the existing lithium battery chemistries, such as Li-ion, Li-$O_2$ or Li-S batteries.

In summary, we firstly discover the MRR/MOR between $Mn^{4+}$/$Mn^{2+}$ as a reversible process, similar to ORR/OER, potentially for conversing or storing enormous energy at a fast rate. The MRR/MOR, occurring on graphene/$MnO_2$ electrode, delivers an anomalous capacity up to 4200 mAh/g, a fast kinetics (2.4× $10^{-3}$ $cm^2$/s, three orders higher than ORR), high reversibility up to 100%, and a high solubility (three orders than $O_2$). As an example, we further use MRR/MOR to invent a new multivalent ion battery (MIB), which uses graphene (or graphene/$MnO_2$) cathode, Zn cathode, and the aqueous electrolyte containing $Zn^{2+}$ and $Mn^{2+}$ ions. The MIBs based on MRR/MOR deliver a high energy density up to 1200 Wh/Kg (several times than lithium ion batteries used in Tesla Model S), a fast charge ability (within 3 minutes, ten times of lithium batteries), and a long cycle life (10000 cycles). Our results indicate that MRR/MOR can be a basic process for a new class of energy conversion or storage systems and the new safety and ecofriendly MIBs open possibility that the electric vehicles run much more miles at one charge. We anticipate that the systems based on MRR/MOR give the society a different choice from lithium and hydrogen economics.

the rechargeable zinc ion battery. *Angew. Chem. Int. Ed.* 51, 933-935 (2012).